# Nanobody interaction unveils structure, dynamics and proteotoxicity of the Finnish-type amyloidogenic gelsolin variant

Toni Giorgino[1,2], Davide Mattioni[1,3,§], Amal Hassan[2,§], Mario Milani[1,2], Eloise Mastrangelo[1,2], Alberto Barbiroli[4], Adriaan Verhelle[5], Jan Gettemans[6], Maria Monica Barzago[3], Luisa Diomede[3] and Matteo de Rosa[1,2,*]

[1] Istituto di Biofisica, Consiglio Nazionale delle Ricerche, Milano, Italy; [2] Dipartimento di Bioscienze, Università degli Studi di Milano, Milano, Italy; [3] Department of Molecular Biochemistry and Pharmacology, Istituto di Ricerche Farmacologiche Mario Negri IRCCS, 20156 Milan, Italy; [4] Dipartimento di Scienze per gli Alimenti, la Nutrizione e l'Ambiente, Università degli Studi di Milano, Milano, Italy; [5] Department of Molecular Medicine, Department of Molecular and Cellular Neuroscience, Dorris Neuroscience Center, The Scripps Research Institute, La Jolla, CA 92037, USA; [6] Nanobody Lab, Department of Biochemistry, Faculty of Medicine and Health Sciences, Ghent University, Ghent, Belgium.

* To whom correspondence should be addressed: Istituto di Biofisica CNR, ℅ Dip di Bioscienze, via Celoria 26, 20133 Milano, Italy. E-mail: teo.derosa@gmail.com

§ These authors contributed equally to this work.

# Highlights

- D187N gelsolin variant is responsible for the most common form of AGel amyloidosis
- We obtained the crystal structure of the second domain of D187N in complex with a nanobody
- D187N substitution increases the conformational flexibility of the protein
- Nanobody binding shifts D187N towards a native-like conformation
- The same nanobody binds to and protects other gelsolin pathological variants (N184K and G167R)
- The nanobody protects from the toxicity induced by gelsolin mutants in *C. elegans*

# Abstract

AGel amyloidosis, formerly known as familial amyloidosis of the Finnish-type, is caused by pathological aggregation of proteolytic fragments of plasma gelsolin. So far, four mutations in the gelsolin gene have been reported as responsible for the disease. Although D187N is the first identified variant and the best characterized, its structure has been hitherto elusive. Exploiting a recently-developed nanobody targeting gelsolin, we were able to stabilize the G2 domain of the D187N protein and obtained, for the first time, its high-resolution crystal structure. In the nanobody-stabilized conformation, the main effect of the D187N substitution is the impairment of the calcium binding capability, leading to a destabilization of the C-terminal tail of G2. However, molecular dynamics simulations show that in the absence of the nanobody, D187N-mutated G2 further misfolds, ultimately exposing its hydrophobic core and the furin cleavage site. The nanobody's protective effect is based on the enhancement of the thermodynamic stability of different G2 mutants (D187N, G167R and N184K). In particular, the nanobody reduces the flexibility of dynamic stretches, and most notably decreases the conformational entropy of the C-terminal tail, otherwise stabilized by the presence of the $Ca^{2+}$ ion. A *Caenorhabditis elegans*-based assay was also applied to quantify the proteotoxic potential of the mutants and determine whether nanobody stabilization translates into a biologically relevant effect. Successful protection from G2 toxicity *in vivo* points to the use of *C. elegans* as a tool for investigating the mechanisms underlying AGel amyloidosis and rapidly screen new therapeutics.

# Keywords

# 1. Introduction

AGel amyloidosis (AGel) is a neglected disease caused by deposition of gelsolin (GSN) amyloids and described for the first time in Finland in 1969 [1]. For a long time, AGel has been associated with the substitution of a single residue of the protein, D187 in the protein second domain (G2) (Figure 1), to either N or Y [2,3]. It has also been considered an endemic pathology in Finland and named Familial Amyloidosis, Finnish-type (FAF). Nowadays, AGel is the preferred name for this disease [4], as new cases have been gradually reported from many other countries demonstrating its worldwide occurrence.

In the last five years, the broader clinical use of genetic tests and the raised awareness of this class of diseases led to the identification of three new AGel forms. A new classification of the disease into three different types, according to the GSN sequence and the organ(s) involved in amyloid deposition, has been therefore proposed and includes: i) a systemic form, caused by D187N and D187Y mutations, respectively known as the Finnish- and Danish-variant; ii) a kidney localized form, associated with the deposition of GSN, either full-length or as fragments, carrying N184K or G167R mutation [5–7]; and iii) a sporadic form, caused by wild-type (WT) GSN deposits surrounding a sellar glioma of the hypophysis [8]. All AGel types share the lack of effective pharmacological therapies that cure the disease targeting the source of toxicity, rather than acting only as palliative, symptomatic treatments.

Among the listed GSN variants, the D187N protein is the best biochemically and biophysically characterized. More than 20 years of *in vitro* and *in vivo* studies led to a consensus on the pathological mechanism underlying Finnish AGel type, although this model has been questioned by recent findings [9]. According to this model, aspartic acid 187 is part of a cluster of residues in the G2 domain of GSN, able to chelate a calcium ion [10]. Its N or Y substitution compromises calcium binding [11–14], leading to the exposure of an otherwise buried sequence, which is recognized by the furin protease [15]. In the Golgi, this intracellular enzyme cleaves GSN producing a C-terminal 68 kDa fragment (C68). C68 is later exported to the extracellular space where it is further processed by matrix metalloproteases, eventually producing 5 and 8 kDa highly amyloidogenic peptides [16]. These fragments rapidly aggregate and deposit in different tissues and organs [17,18]. In stark contrast to the extensive biochemical knowledge available on the D187N mutant, its crystal structure has never been obtained, limiting the mechanistic understanding of GSN instability and aberrant proteolysis.

This study aims at characterizing the crystal structure of the isolated G2 domain (Figure 1) of the D187N protein (D187N$_{G2}$) by exploiting a recently-developed nanobody (Nb) targeting GSN [19]. Different Nbs able to bind mutated GSN and to detect or prevent its aggregation, have been developed and tested[19–22]. Among them, Nb11 proved to be the most efficient one. Studies performed *in vitro* and *in vivo* demonstrated that Nb11 binds G2 domain of GSN with high affinity, irrespective of calcium, and protects the mutated domain from furin proteolysis, thus skipping the first event of the aberrant proteolytic cascade (19–22). Inspired by the recent use of Nbs as an unique tool for structural biological studies [23], we employed Nb11 to increase the stability of D187N$_{G2}$.

The successful co-crystallization of D187N$_{G2}$ in complex with Nb11 (D187N$_{G2}$:Nb11) showed that the nanobody protects D187N$_{G2}$ from furin-induced proteolysis, stabilizing the G2 C-terminal linker. Such stabilization is achieved allosterically since the Nb11 binding site locates far from the furin cleavage site. We complemented the structural results cross-referencing molecular dynamics (MD) simulations insights with thermal denaturation studies and furin proteolysis assays. These studies were extended to other mutations causing AGel, such as G167R and N184K.

In the absence of cellular or animal models recapitulating G167R and N184K-related AGel as well as the toxicity of the WT or mutated G2 domains, we decided to employ the invertebrate nematode *Caenorhabditis elegans* as "biosensor", able to recognize proteins which exert *in vivo* a biologically relevant effect [24–27]. This approach takes advantage of the ability of the pharynx of worms, fundamental for their feeding and survival, to be inhibited when it meets molecules acting as chemical stressors [28]. This nematode-based method has been widely applied to recognize the toxicity of different amyloidogenic proteins *in vivo,* demonstrating that singular molecular mechanisms underlie their proteotoxic activity [24–27,29]. The protein folding, oligomerization propensity and the exposure of hydrophobic residues on the outside of the protein are relevant for the toxic action of β-amyloid (Aβ) and HIV-matrix protein p17 [25,27,29]. Instead, amyloidogenic cardiotoxic light chains are recognized as stressors by *C. elegans* thanks to their ability to interact with metal ions and continuously generate reactive oxygen species [24,26].

Our findings indicate that *C. elegans* efficiently recognizes the proteotoxic potential of the G2 domains and can discriminate between different level of toxicity. Furthermore, the stabilizing effects induced by Nb11 on G2 translated into an effective protection *in vivo.* These observations point to the use of this nematode-based model as a valuable tool for investigating the mechanisms underlying AGel.

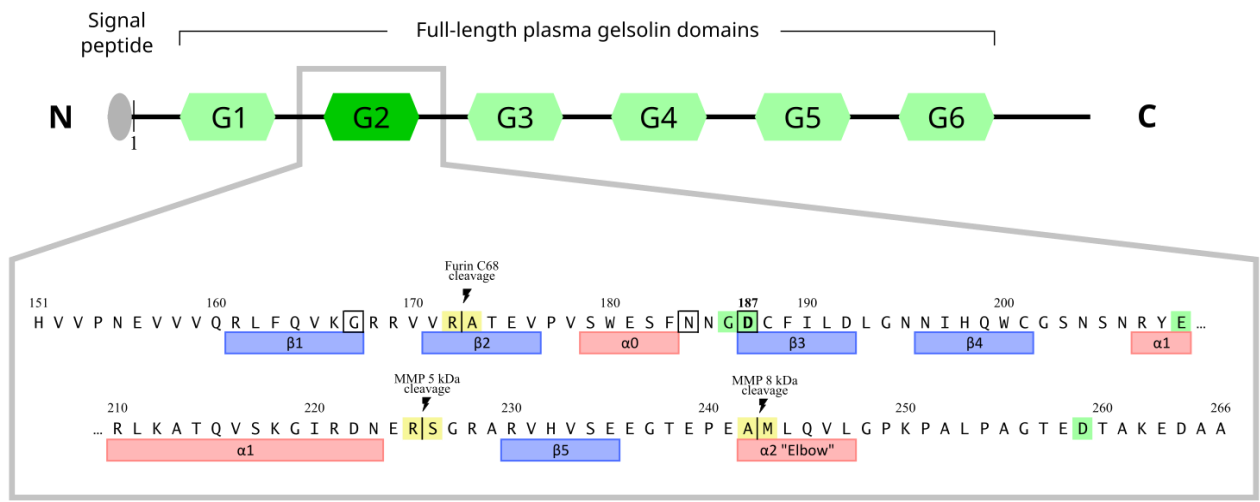

*Figure 1: Domain architecture of plasma gelsolin and sequence of the G2 domain. The G2 construct comprises residues 151 to 266 around the second domain of full-length plasma gelsolin (residue numbering excludes the signal peptide). Secondary structure elements (α0 to α2, β1 to β5) are indicated below the sequence; calcium-coordinating residues are marked in green, and known amyloidosis-causing amino-acid mutations (D187N object of this study, as well as D187Y, G167R, N184K) are boxed. Vertical bars indicate the locations of aberrant processing by furin and matrix metallo-proteases (MMP), ultimately leading to fibril formation and AGel.*

# 2. Materials and methods

## 2.1 Protein production

If not otherwise stated, all chemicals are from Sigma-Aldrich (Merck Millipore Ltd., Tullagreen, Carrigtwohill, Co. Cork, IRL) and of the best available purity. All preparative and analytic chromatographies were performed on a ÄKTA pure 25 system (GE Healthcare, Uppsala, Sweden) using prepacked columns from the same company.

### 2.1.1 Gelsolin expression and purification

Constructs and expression conditions for GSN variants as isolated G2 domain or full-length proteins were performed as already reported [30,31]. The here named G2 refers to the shorter G2s construct in [30] and spans residues 151-266 of the GSN, according to the mature plasma isoform (Figure 1). Purification followed the protocol previously described [30]. ΔG2 construct for the expression of $WT_{\Delta G2}$ harbours a WT sequence but lacks the first beta-strand (β1); it spans residues 168-266 and it is purified following the same protocol developed for the other G2 domains [31]. Briefly, G2 domains were passed through a Ni-chelating column (HisTrap) and gel filtered in a Superdex 75 following the removal of the His-tag. Full-length proteins went through three chromatographic steps without cleaving the N-terminal tag, HisTrap, MonoQ and gel filtration on a Superdex 200. G2 domains were stored in a 20 mM HEPES solution, pH 7.4, containing 100 mM NaCl and 1 mM $CaCl_2$ whereas full-length GSN in 20 mM HEPES solution, pH 7.4, containing 100 mM NaCl, 1 mM EDTA and 1 mM EGTA. Whenever required proteins were concentrated or buffer-exchanged in centrifugal filters Amicon® Ultra (Merck Millipore Ltd., Tullagreen, Carrigtwohill, Co. Cork, IRL) with cutoffs of either 10k or 30k.

### 2.1.2 Nb11 expression and purification

The synthetic gene coding for Nb11 [19] fused to the thrombin cleavage site and 6xHis tag at the C-terminus was purchased by Eurofins genomics (Ebersberg, Germany). The sequence was optimized for *E. coli* codon usage and the gene cloned in a pET11 vector (Merck Millipore Ltd., Tullagreen, Carrigtwohill, Co. Cork, IRL). Tagged Nb11 was produced in SHuffle® T7 *E. coli* cells (New England BioLabs Inc., Ipswich, USA) grown in LB medium. Once OD at 600 nm reached the value of 0.6, the expression of the gene was induced with 0.5 mM Isopropyl β-D-1-thiogalactopyranoside and cells harvested by centrifugation 16 h after incubation at 20 °C. Cells were resuspended in 20 mM $Na_2PO_4$, pH 7.4, containing 500 mM NaCl (supplemented with DNase I and cOmplete™ Protease Inhibitor Cocktail), lysed with a Basic Z Bench-top cell disruptor (Constant system Ltd., UK) operating at 25 kPsi and centrifuged at 38,000 RCF. The clarified crude extract was loaded on a 5 ml HisTrap and Nb11 eluted stepwise with the lysis buffer supplemented with 500 mM imidazole. Fractions enriched in Nb11 were passed through a Superdex 75, equilibrated with 20 mM HEPES, pH 7.4, 100 mM NaCl.

## 2.2 Crystallization, structure solution and analysis

### 2.2.1 Crystallization

$D187N_{G2}$:Nb11 complex for crystallization experiments was prepared by mixing equimolar amounts of the individual protein. The complex was loaded on a Superdex 75 increase (equilibrated with 20 mM HEPES, 100 mM NaCl, 1 mM $CaCl_2$, pH 7.4), obtaining a single peak, consistent with the theoretical molecular weight of the complex, which was concentrated to 12 mg/ml in the same buffer. This sample was used for extensive crystallization screening using an Oryx-8 crystallization robot (Douglas Instruments Ltd, UK) and several commercial solutions in a sitting-drop set up. The purified complex (0.15/0.25 μl) was mixed with 0.25/0.15 μl of the reservoir solutions. Two different conditions yielded crystals after 2-6 days at 20 °C, namely (i) 0.1 M potassium thiocyanate, 30 % poly(ethylene glycol) methyl ether 2000 and 4.0 M sodium formate and (ii) 0.1 M potassium phosphate, pH 6.2, 10% (v/v) glycerol and 25% v/v 1,2 propanediol. Crystals were soaked with the respective reservoir solution supplemented with 20% glycerol, flash-frozen in liquid $N_2$ and diffraction data were collected at beamline ID23-1 (European Synchrotron Radiation Facility, Grenoble, France).

### 2.2.2 Data processing, structure solution, refinement and analysis

The two datasets (orthorhombic and tetragonal crystals grown in condition (i) and (ii), respectively) were processed with XDS [32], scaled with Aimless [33], and the structures solved by molecular replacement using the program phaser [34] and the PDB: 4S10 as the searching model. The orthorhombic structure (1.9 Å) was refined with *phenix.refine* [35] and manual model building was performed with COOT [36]. The final coordinates were deposited in the RCSB database with accession code PDB: 6H1F. The tetragonal structure (2.4 Å) could be refined only partially due to the poor quality of the diffraction (see Table 1 for the complete data collection and refinement statistics). The structural analysis was performed using the orthorhombic structure, except where indicated; all of the figures were prepared with either PyMOL or VMD [37,38].

All the other structures used for the B factor analysis were subjected to 3 cycles of refinement with *phenix.refine*. B factors for the $WT_{G2}$:Nb11 and $D187N_{G2}$:Nb11 were averaged between asymmetric molecules or between datasets, respectively. The rest of the analysis was performed as reported earlier [39].

## 2.3 Molecular dynamics (MD) simulations

The crystal structure of G2 domain in complex with Nb11 (PDB: 4S10) was used to build the initial configuration for MD runs. The structure editing and building facilities provided by the HTMD software package [40] were used to construct a set of systems by selecting either the G2 domain only (chain D of the original PDB file) or the domain together with the bound Nb11 (chains D and B). *In silico* mutations were applied to restore residues 226 and 228 to their WT amino acids, to generate the D187N and N184K systems, and to remove the coordinated $Ca^{2+}$ ion where appropriate. The titratable side chains and initial H bond network were then optimized at pH 6.5 via the *proteinPrepare* procedure of HTMD [41]; systems were solvated with transferable intermolecular potential with 3 points (TIP3P) water at 100 mM NaCl ionic strength and parameterized with the CHARMM36 force field [42]. The systems thus prepared were minimized and equilibrated for 4 ns in constant pressure conditions at 1 atm, yielding orthorhombic boxes with a water buffer of at least 15 Å per side. All simulations were conducted via the ACEMD software with a timestep of 4 fs, particle-mesh Ewalds long-range electrostatics treatment and the hydrogen mass repartitioning scheme [43].

Each system was set for a production run in the number volume temperature (NVT) ensemble. Simulations of the G2 system without Nb11 were completely unrestrained and therefore the globular domain was free to diffuse in the solvent. To obtain a more efficient simulation box for the G2:Nb11 elongated complex while still preventing self-interactions with periodic images, its rotational diffusion was restricted by restraining Cα atoms of the Nb11's secondary structure elements with a harmonic force of 0.025 kcal/mol/Å$^2$. No restraint was applied to the G2 domain, nor to the contact region of Nb11. Runs were interrupted at 800 ns or when the full solvation of the C-terminus caused its extension outside of the simulation box. The $WT_{G2}$:Nb11 simulation was also truncated at 750 ns because the complex became transiently unbound. Local conformational flexibility was assessed computing the root-mean-square fluctuation (RMSF) of backbone atoms, aggregated by residue; values are reported as the equivalent B factors according to the equation $B=8\pi^2/3\ RMSF^2$.

## 2.4 Thermal stability

### 2.4.1 Circular Dichroism (CD) spectroscopy

CD measurements were performed with a J-810 spectropolarimeter (JASCO Corp., Tokyo, Japan) equipped with a Peltier system for temperature control. All measurements were performed on 15 µM G2, Nb11, or the complex in 20 mM HEPES solution containing 100 mM NaCl and 1 mM $CaCl_2$ at pH 7.4. Temperature ramps were recorded from 10°C to 95°C (temperature slope 50 °C/h) in a 0.1 cm path length cuvette and monitored at 218 nm wavelength.

### 2.4.2 Thermofluor

Thermodynamic stabilities were also evaluated in the presence of Sypro Orange, a fluorogenic probe unspecifically binding hydrophobic surfaces and with excitation/emission spectra compatible with standard qPCR machines. WT and mutated G2 domains with or without equimolar Nb11, were diluted to 1 mg/ml in 20 mM HEPES solution containing 100 mM NaCl and 1 mM $CaCl_2$, pH 7.4. Each solution was mixed with 3 µl of a 1/500 (v/v) dilution of Sypro Orange , in a total volume of 20 µl. Fifteen µl of each sample were transferred to multiplate® PCR Plates, sealed with Microseal® 'B' Film, and analysed in triplicate in an MJ Mini™ Thermal Cycler (hardware and consumables from Bio-Rad Laboratories Inc., Hercules, USA). The temperature was increased from 10 °C to 100 °C in 0.2 °C steps with 10 s equilibration before each measurement. Fluorescence intensity was measured within the excitation and emission wavelength ranges of 470–505 and 540–700 nm, respectively. $T_m$ was calculated as the minimum of the first-derivative of the traces using the manufacturer software and the value is reported as the average of triplicate measures.

## 2.5 Furin assay

Furin cleavage assays were performed in a total volume of 30 µl, using 1 U of commercial furin enzyme (New England BioLabs Inc., Ipswich, Massachusetts, USA) and 1 mg/ml of full-length WT and mutated GSN in 20 mM 2-(N-morpholino)ethanesulfonic acid, pH 6.5, containing 100 mM NaCl and 1 mM $CaCl_2$ in the presence or absence of 1 mg/ml of Nb11 (roughly 1:6 GSN:Nb11 molar ratio). To monitor the susceptibility to proteolysis, 12 µl aliquots of the reaction mix were collected right upon addition of furin and 3 h after incubation at 37 °C. The reaction was blocked by adding to each sample 4 µl of Sodium Dodecyl Sulphate (SDS) loading buffer 4X (BioIO-RadAD Laboratories Inc., Hercules, USA) supplemented with 0.7 M β-mercaptoethanol and by incubation at 90 °C for 3 min. Proteolysis reaction was monitored by SDS - PolyAcrylamide Gel Electrophoresis using ExpressPlus™ PAGE (12%) and the provided running buffer (GenScript Biotech Corp., USA ).

## 2.6 Proteotoxicity studies on *C. elegans*

Bristol N2 strain was obtained from the *Caenorhabditis elegans* Genetic Center (CGC, University of Minnesota, Minneapolis, MN, USA) and propagated at 20°C on solid Nematode Growth Medium (NGM) seeded with *E. coli* OP50 (CGC) for food. The effect of G2 domains on pharyngeal behavior was evaluated as already described [27]. Briefly, worms were incubated with 1-1000 µg/ml of G2 domains (100 worms/100 µl) in 2 mM HEPES solution containing 1 mM NaCl and 0.1 mM $CaCl_2$, pH 7.4. Equimolar concentrations of $WT_{G2}$ or $WT_{\Delta G2}$ (18 µM, corresponding to 250 µg/ml of WTG2) were administered to worms in the same conditions. Hydrogen peroxide (1 mM) was administered in dark conditions as a positive control. After 2 h of incubation on orbital shaking, worms were transferred onto NGM plates seeded with OP50 *E. coli*. The pharyngeal pumping rate, measured by counting the number of times the terminal bulb of the pharynx contracted over a 1-minute interval, was scored 2 and 24 h later. Control worms were fed 2 mM HEPES solution containing 1 mM NaCl and 0.1 mM $CaCl_2$, pH 7.4 (Vehicle) only. To evaluate the protective effect of Nb11, worms were fed for 2 h G2 domains alone (250 µg/ml for $WT_{G2}$ and $D187N_{G2}$ and 100 µg/ml for $N184K_{G2}$ and $G167R_{G2}$, corresponding to 19 and 8 µM, respectively), 8-19 µM Nb11 alone, or G2 domains previously pre-incubated for 10 min at room temperature under shaking conditions with equimolar concentration of Nb11 to allow the formation of the complex. Nematodes were then transferred to NGM plates seeded with fresh OP50 *E. coli* and the pumping rate was scored after 2 and 24 h. Worms were also exposed to Vehicle in the same experimental conditions.

# 3. Results

## 3.1 Crystal structure of the D187N$_{G2}$:Nb11 complex

Crystals of the isolated D187N$_{G2}$:Nb11 complex readily appeared in different conditions, allowing the collection of two X-ray diffraction datasets of different quality. Particularly, the crystal grown in condition (i) above, belonging to the P2$_1$2$_1$2$_1$ space group, diffracted to 1.9 Å resolution, and the one that appeared in condition (ii), belonging to the P4$_1$2$_1$2 space group, diffracted to 2.4 Å. The resulting models are hereafter referred to as the *orthorhombic* and *tetragonal* structure, respectively (Figure 2); the complete list of the data collection and refinement statistics is reported in Table 1. Due to the difference in quality of the data, the *orthorhombic* structure is used to infer the impact of the D187N mutation on the GSN structure (Figure 2), unless otherwise stated.

|  | **Orthorhombic** | **Tetragonal** |
|---|---|---|
| PDB ID | 6H1F | — |
| **Data collection** | | |
| Space group | P 2$_1$ 2$_1$ 2$_1$ | P 4$_1$ 2$_1$ 2 |
| Cell dimension: a, b, c (Å) | 33.8, 46.8, 132.1 | 77.0, 77.0, 88.8 |
| Unique reflections | 17,289 | 10,995 |
| Resolution range (Å) | 46.8-1.9 (1.94-1.90) | 46.4-2.4 (2.49-2.40) |
| CC$_{1/2}$ | 0.990 (0.714) | 0.999 (0.800) |
| Completeness (%) | 99.8 (99.9) | 100 (100) |
| Multiplicity | 5.1 | 16.9 |
| **Refinement** | | |
| Resolution range (Å) | 44.1-1.9 | 46.4-2.4 |
| R$_{work}$ / R$_{free}$ (%) * | 19.9/23.3 | 18.6/26.0 |
| RMSD | | |
|    Bonds (Å) | 0.006 | 0.008 |
|    Angles (°) | 0.786 | 0.964 |
| Ramachandran plot | | |
|    In preferred regions (%) | 96.0 | 95.7 |
|    Outliers (%) | 0.0 | 0.4 |
| B-factors (Å$^2$) $^§$ | 30 | 66 |

*Table 1: Data collection and refinement statistics.* Notes: Values in parentheses refer to the highest resolution shell. * $R_{work} = \Sigma hkl||Fo| - |Fc||/\Sigma hkl|Fo|$ for all data, except 10%, which were used for $R_{free}$ calculation. $^§$ Average temperature factors over the whole structure.

Both the structures obtained confirm that Nb11 binds the G2 domain over an extended area including the β5-loop-α2 region (residues 230-234, 238-245) and the loop-β4 stretch (residues 193-198) as already observed for WT gelsolin (PDB: 4S10, [19]). The binding region is opposite from the aberrant cleavage site (residues 168-172), which had previously raised suspicions of a crystallographic artifact. The structures obtained here rule out this hypothesis, since it is unlikely that three different crystal packings (WT$_{G2}$:Nb11 belongs to P1 space group) artificially stabilize the same assembly (Figure 2A). Thus, the 3D structures of G2:Nb11 appear unable to explain how the binding to Nb11 in a distal area could *shield* G2 from furin proteolysis without additional data.

The second remarkable feature of the WT$_{G2}$ and D187N$_{G2}$ assemblies is that the G2 structures are almost identical (root mean squared deviation, RMSD, 0.25 Å over 90 Cα atoms). Besides, two striking differences were observed: the absence of the coordinated calcium ion, and a shorter stretch of ordered C-terminal segment in D187N$_{G2}$ (Figure 2A). In the C-terminus of mutated G2, we could only model up to residue 258 and 255 in the *orthorhombic* and *tetragonal* structure, respectively (Figure 2A), suggesting that the remaining stretch of the protein is highly flexible and invisible to X-rays (the last resolved amino acid in WT$_{G2}$ is 261). The disorder prediction algorithm MobiDB-lite [44] indeed reports a putative disordered region around residues 220-258 owing to its poly-ampholitic character (Supplementary Figure S1). Besides the C-terminus, all residues are modelled but a few side chains of the Nb11 chain (namely V2, K65 and L128).

The superimposition of the *orthorhombic* structure of D187N$_{G2}$ to that of the WT$_{G2}$ (PDB: 1KCQ), shows no major conformational changes (as is the case for the G167R variant) nor the rearrangement of the hydrogen bond network in the core of the domain in the N184K variant, unlike it was observed for the other G2 mutants (Figure 2B). Residues G186, D187 and E209 of the calcium-coordinating cluster show remarkable conformational conservation (D259, the fourth residue of the cluster, lies in the unresolved C-terminal stretch).

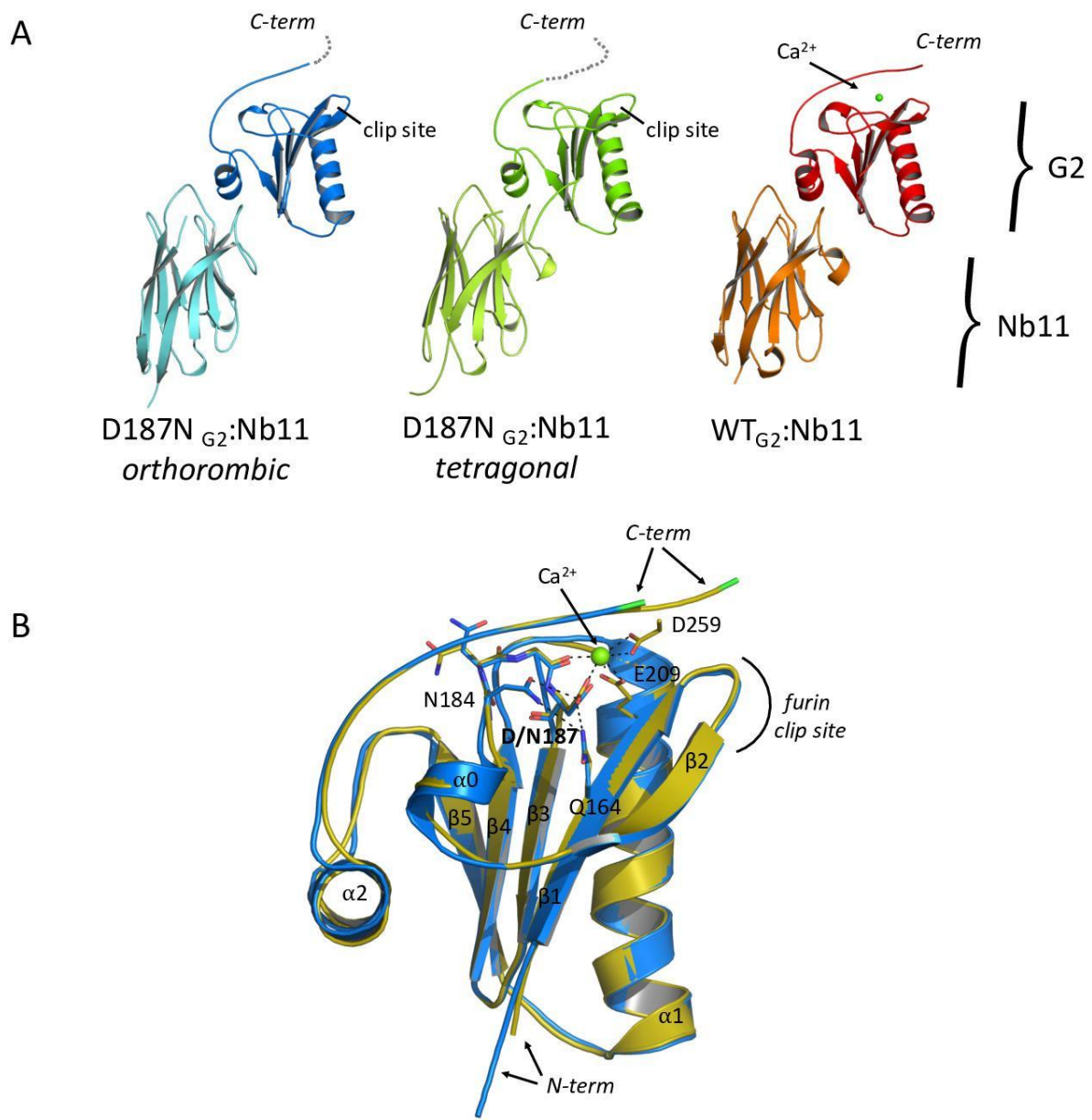

*Figure 2: Crystal Structure of the complexes formed by WT or D187N Gelsolin G2 domains with the Nb11 nanobody. A) The overall structure of the D187N$_{G2}$:Nb11 complexes compared to the WT$_{G2}$:Nb11. Complexes are represented in cartoon and colored blue/cyan (orthorhombic D187N$_{G2}$:Nb11), green/light green (tetragonal D187N$_{G2}$:Nb11) and red/orange (PDB:4S10, WT$_{G2}$:Nb11); the calcium ion is shown as a green sphere. The C-terminal residues unresolved in the D187N structures are arbitrarily displayed by a dashed grey line. B) Superimposition of the D187N$_{G2}$ domain extrapolated from the orthorhombic structure with the WT$_{G2}$ (PDB:1KCQ). Residues of the calcium cluster (green sphere) and those found altered in the N184K structure are shown as sticks. WT and D187N variants share high structural similarity (RMSD: 0.25 Å over 90 Cα atoms).*

## 3.2 Impact of the D187N mutation on the structure of the G2 domain

A large body of evidence indicates that the presence of D187N mutation mainly affects the ability to bind the calcium ion [11–14]. In $WT_{G2}$ the calcium binding site is a highly polar cavity delimited by the protein core, the side chain of K166 and the C-terminal tail (Figure 3A, left). The ion is pentacoordinated by the side-chains of residue D187, E209 and D259 and by the main-chain O atom of G186. The site is open to the solvent on the side opposite to K166, where two ordered water molecules are visible in the $WT_{G2}$ structure (Figure 3A, left).

In the *orthorhombic* structure of $D187N_{G2}$, although the overall geometry of the calcium binding is maintained, some differences are visible. The residue D259 could not be modeled and the presence of a water molecule in its place (bound to the O atom of T257) shows that the residue is extruded from the cavity. Two more water molecules are found overlapping neither with the calcium ion nor with other solvent molecules present in the $WT_{G2}$ structure. The side-chain of K166 also undergoes small rearrangements: its amino tip becomes more flexible, based on the quality of the electron density, and the interaction with residues 187 and 259 is either lost or weakened. The N187-K166 distance is 4.6 Å, whereas 3.1 Å separate D187 from K166 in the $WT_{G2}$ due to a stronger polar interaction (Figure 3A, middle). This small reorganization of the calcium binding residues, as well as the lack of the coordinated ion, are sufficient to create a small cavity (Figure 3A, right, computed by PyMOL [37]), a known source of instability in proteins [45]. Even though this cleft is solvent-exposed and accessible, the high concentration of its acidic residues might be the force driving the stretching of both D259 residue and the whole C-terminal tail.

The same cavity is fully open in the *tetragonal* structure of $D187N_{G2}$, where we were able to model only up to residue 255. The electron density for the calcium site is of poor quality (data not shown), the side-chain of K166 is barely visible and no solvent molecule could be modelled. The *orthorhombic* and *tetragonal* structures seem to represent two intermediate states toward the domain's (partial) unfolding.

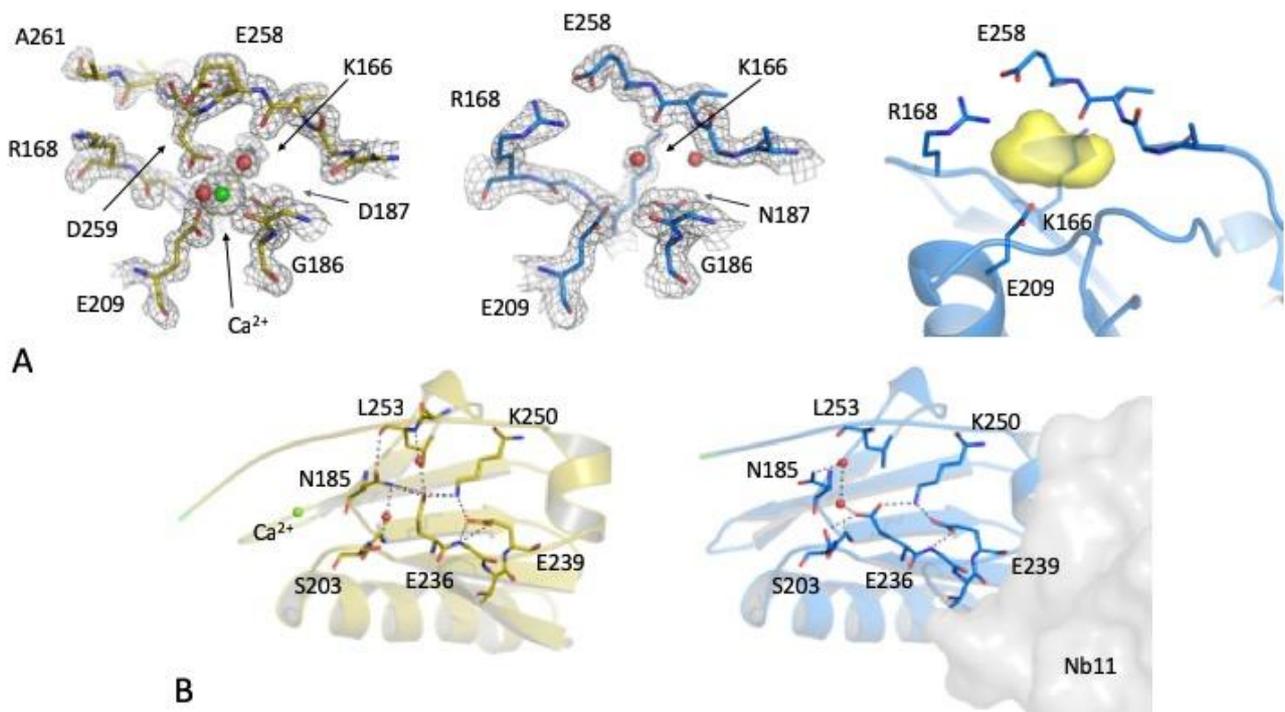

*Figure 3: Impact of the D187N mutation on the structure of the gelsolin G2 domain. A) Details of the calcium binding site in the $WT_{G2}$ (yellow, PDB: 1KCQ) and $D187N_{G2}$ (blue). Electron density is contoured at 1.5 σ. The cavity volume induced by D187N substitution was generated by PyMOL and is shown in yellow. B) Reorganization of the H-bond network in the C-terminal "arm" of the domain of the $D187N_{G2}$ variant (blue cartoon and sticks) in comparison with the $WT_{G2}$ (yellow). In 1KCQ cadmium substitutes the physiological ion; here it is labelled as calcium for readability ($Ca^{2+}$, green sphere); ordered water molecules playing a structural role in the respective regions are represented as red spheres*

## 3.3 Nb11 binding modulates C-terminal disorder allosterically

Based on our structures, one can hypothesize that Nb11 binding induces $D187N_{G2}$ in a proteolysis-resistant conformation, but crystallographic analysis alone is not sufficient to assess whether the flexibility of the $D187N_{G2}$ C-terminal tail and exposure of the calcium cavity lead to its susceptibility to furin proteolysis. To investigate the impact of the mutation on local kinetics or a possible conformational selection effect by Nb11 binding, we analyzed the domain's dynamics via MD simulations. We simulated the $WT_{G2}$ and $D187N_{G2}$ systems in the presence or absence of bound Nb11 (Table 2), starting from the crystallographic poses reported in PDB: 4S10 (see Methods section).

| Sample | Nb11 | $Ca^{2+}$ | Simulated time (ns) | C-terminal disorder onset |
|---|---|---|---|---|
| **$WT_{G2}$** | - | + | 800 | Not observed |
| **$WT_{G2}$** | + | + | 750 | Not observed |
| **$D187N_{G2}$** | - | - | 748 | After 83 ns |
| **$D187N_{G2}$** | + | - | 512 | After 40 ns |

*Table 2: Summary of the molecular dynamics runs. Simulated time and behavior of the C-terminal stretch are reported. "C-terminal disorder onset" indicates the time of onset of C-terminal segment mobility (residue 248 and beyond), defined as R186-T260 distance > 20 Å. Afterwards, the C terminal segment ("arm") is solvated and strongly fluctuating.*

The most striking result was the fast opening and solvent-exposure of the C-terminus segment, which occurred within the first hundred nanoseconds, in all of the structures of $D187N_{G2}$ lacking the $Ca^{2+}$ ion (Table 2). In the crystallographic structures of the $WT_{G2}$ protein, the C-terminal region up until residue 259 lies near the β1-β2 hinge loop which hosts the furin cleavage site ($R_{169}$-V-V-$R_{172}$), indicating a possible steric protection mechanism already suggested by Huff et al. [14]. In consistency with our crystallographic results, MD did not uncover further major "static" structural rearrangements between the $WT_{G2}$, $D187N_{G2}$ and $D187N_{G2}$:Nb11 in the timescales tested.

Local (residue-wise) fluctuations across the G2 residues (Figure 4A, red curve) indicate two major areas of destabilization in the $D187N_{G2}$ with respect to the $WT_{G2}$, namely (a) the region 179-183 (helix α0) is located; and (b) the *arm* of the domain, i.e. the C-terminal region up to residue 240 including the β5-α2 loop, the α2 helix and the opening C-terminal arm plus its *elbow* at helix α2. Binding of Nb11 to the $D187N_{G2}$ (Figure 4A, cyan curve) almost entirely abolished the fluctuations in both regions. In other words, Nb11 essentially restores the fluctuation dynamics of the mutant to the low level present in the $WT_{G2}$. The simulation of the $WT_{G2}$:Nb11 complex confirmed that the Nb11 does not significantly alter the local stability of the $WT_{G2}$, although it partially stabilizes the β5-α2 region where it binds.

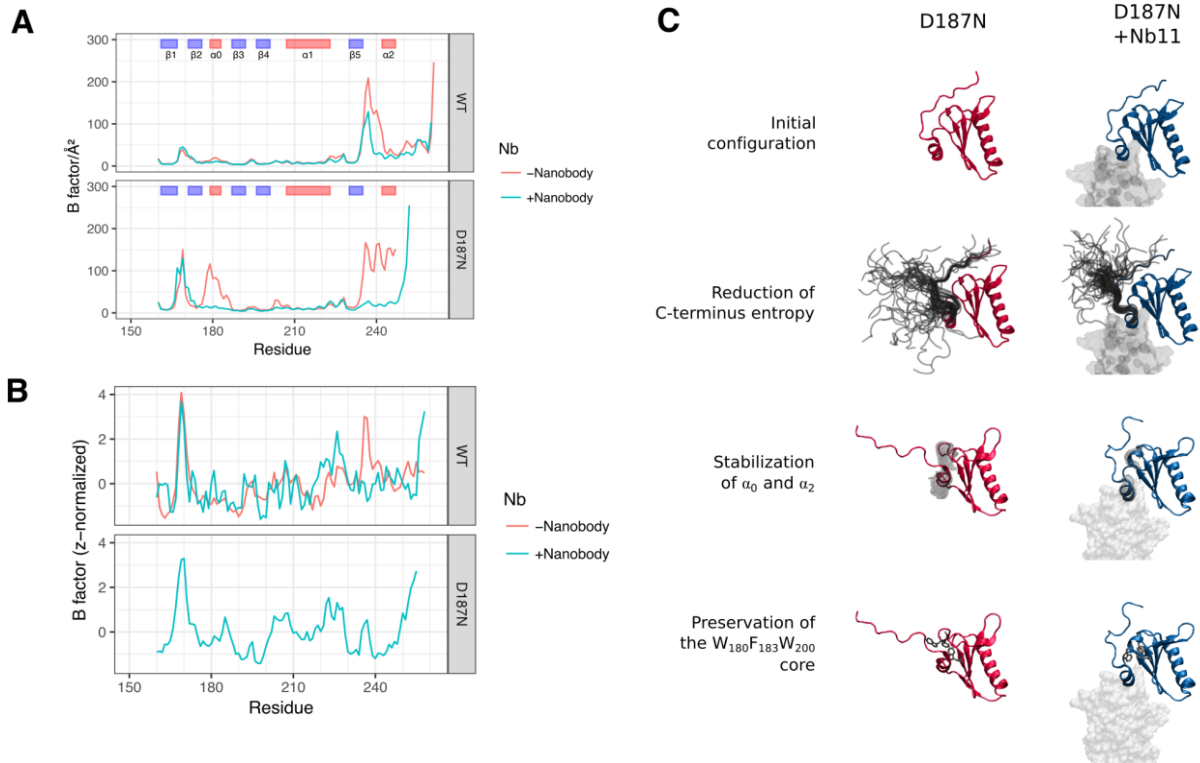

*Figure 4: Dynamics and fluctuations of the $D187N_{G2}$ variant in the presence or absence of Nb11. A)* Plot of the static B factors per residue calculated by the simulation of the $WT_{G2}$ and $D187N_{G2}$ in the presence (cyan line) or absence (red) of Nb11. Structured regions are indicated in blue (β sheets from β1 to β5) and pink (α helices α0 to α2). Furin cleavage site is in the β1-β2 loop. *B)* Normalized B factors ($B_{z-score}$) extracted by the crystallographic structures of $WT_{G2}$, $WT_{G2}$:Nb11 and $D187N_{G2}$:Nb11, color-coded as in panel A. *C)* Overview of the main molecular features and conformational diversity observed during the simulations of $D187N_{G2}$ in the presence or absence of Nb11 (light grey). Dark grey: snapshots of the disordered C-terminus from the conformational ensemble.

The main molecular events underlying the destabilization induced by the D187N mutation are summarized in Figure 4C. In the absence of Nb11, the fluctuations in $D187N_{G2}$ lead to the already described flexibility of the C-terminal tail with the increase of its entropy. Then the transient displacement of the α2 helix occurs, followed by the disruption of the hydrophobic core triad composed by W180, F183 and W200 residues (Supplementary Figure S2). Noteworthy, the ConSurf algorithm indicates that F183, belonging to the α0 helix, is strongly conserved in gelsolin-like domains (see Discussion) [46]. Even though binding of Nb11 has only a minor impact on the resolved pose of the C-terminal tail, its conformational entropy was reduced compared with $D187N_{G2}$ alone, thus likely preventing the rearrangement of the α2 helix and the exposure of the core residues.

To validate the computational findings, we further compared the amount of disorder obtained from the simulations with the crystallographic data. Although not ideal to investigate protein dynamics, crystallographic data in the 2 Å resolution range does contain information on the disorder and mobility of the single atoms in the crystal [47]. To this purpose, B factors shown in Figure 4B were extracted from the structures of D187N$_{G2}$:Nb11, WT$_{G2}$ (PDB: 1KCQ) and WT$_{G2}$:Nb11 (PDB: 4S10) (the structure of the D187N$_{G2}$ variant alone is unavailable). B values were averaged between datasets or symmetric molecules (where available) and normalized to zero mean and unit variance in order to obtain a B$_{z\text{-score}}$ comparable between models (Figure 4B). As expected, we observe a sharp decrease of WT$_{G2}$ and D187N$_{G2}$ B$_{z\text{-score}}$ at the Nb11 binding interface, indicating that this region becomes less dynamic. Stabilization also propagates through the C-terminus and the α2 helix (that is partly involved in the interaction with Nb). The hinge loop formed by amino acid residues 167-170, which hosts the sequence recognized by furin, results very dynamic in both WT$_{G2}$ and D187N$_{G2}$ structures (with and without Nb11) suggesting that the presence of the mutation does not directly affect the conformation or the stability of this loop.

The MD analysis identified two instability hotspots in the mutated G2 domain: the α0 helix and the stretch comprising β5-α2 loop and α2 helix. In the first region we were not able to detect any significant differences between the structure of the WT$_{G2}$ and that of D187N$_{G2}$ stabilized by Nb11 (data not shown). As shown in Figure 3B (left panel), the WT$_{G2}$ β5-α2 loop is connected to the C-terminal tail and the calcium binding site through a solvent-mediated H-bond network. The structure of the mutant reveals a reorganization of the polar contacts mostly due to a drift of the backbone in the 235-239 stretch and the different conformation of side-chains of residues 185 and 236 (Figure 3B, right panel). Overall, these observations indicate that the D187N substitution causes a slight loss in connectivity of the G2 domain essentially due to the loss of the coordinated Ca$^{2+}$. Major rearrangements observed in the simulations are thus likely prevented by the binding to Nb11, which partly compensates for the loss of the calcium ion and forces the first half of the C terminus of D187N in a WT-like conformation.

## 3.4 Nb11 protects all pathogenic GSN variants from aberrant furin proteolysis

The ability of Nb11 to bind D187N-mutated GSN and to inhibit its proteolysis by furin has been extensively demonstrated *in vitro* and *in vivo* [19]. Recently, we have characterized two novel GSN pathogenic variants reported as responsible for a renal localized amyloidosis, G167R and N184K [30,31]. Structural and functional analyses revealed that the N184K substitution induces a reorganization of the connection in the core of the G2 domain; contrarily, G167R mutation promotes the dimerization of the protein by a domain swap mechanism. Similarly to D187N, both substitutions impair protein stability and ultimately lead to protein degradation and aggregation. Therefore, we wondered if Nb11 would also be able to bind the variants responsible for the renal disease and protect these proteins against furin proteolysis. The susceptibility to furin proteolysis of full-length WT and the three aforementioned mutants was tested in the presence or the absence of a 6-fold molar excess of Nb11.

Full-length activated (i.e. calcium-bound) GSN is a flexible protein prone to unspecific proteolysis, a phenomenon that is more evident for the mutants (Figure 5, band marked with *). Its susceptibility to furin cleavage was therefore tested over a short incubation time (3 h) at 37 °C. Furin activity produces two fragments, the well-known C68 and one here named N17, which is visible only when proteolysis proceed to a larger extent, as is the case for the G167R variant. Nb11 dramatically reduced the susceptibility to furin of all the tested variants, either abolishing any trace of its activity, or significantly slowing down the process.

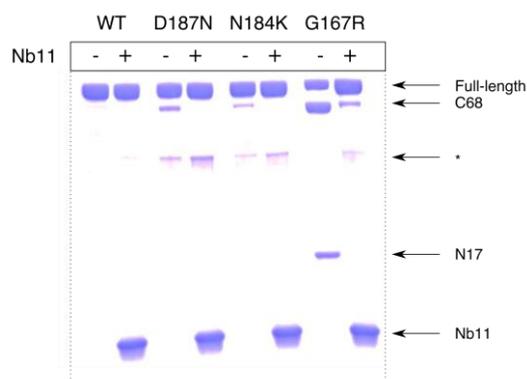

*Figure 5: Effect of Nb11 on the susceptibility of full-length GSN variants to proteolysis. Susceptibility to furin proteolysis of WT, D187N, N184K and G167R variants in their full-length form evaluated in the presence (+) or the absence (-) of Nb11. C68 and N17 are the two fragments produced by furin activity, named after their molecular weight in kDa. \* Indicates unspecific proteolysis product due to contaminants.*

## 3.5 Nb11 stabilizes the mutated G2 fold

The protection exerted by Nb11 on the furin proteolysis cannot be simply explained by steric hindrance. Moreover, our structural and computational analyses pointed at a stability modulation effect. To further elucidate the molecular mechanism underlying the action of the nanobody, we assayed Nb11's impact on the thermal stability of the G2 variants. Denaturation of the proteins was induced by a linear temperature gradient (from 20°C up to 90°C) and studied by CD spectroscopy at 218 nm (sensitive to α-helical content) in the presence of saturating calcium concentration. $WT_{G2}$ and all the mutated domains showed a sigmoidal increase in ellipticity upon unfolding (Supplementary Figure S3). Contrarily, Nb11 signal decreased over time as the immunoglobulin-like domain lacks α-helical content. Interpretation of the melting curves of the complexes is therefore hindered by the opposing behavior of the two proteins.

To analyze CD data, we compared the denaturation curve obtained for the G2:Nb11 complex with the arithmetical sum of the experiments performed on the individual subunits of the complex. We hypothesize that, in the case there is no stabilization, the curve of the complex should roughly superimpose to the curve of the sum, because the G2 domain and Nb11 unfold independently. Indeed, this was the behavior observed for the $WT_{G2}$, in which the binding to Nb11 seems not to have any impact on the protein stability (Figure 6A). In contrast, all the pathological G2 mutants showed significant differences between the denaturation profile of the sum and that of the *experimental* complex (Figure 6A).

|  |  | $WT_{G2}$ | $G167R_{G2}$ | $N184K_{G2}$ | $D187N_{G2}$ |
|---|---|---|---|---|---|
| **- Nb11** | $T_m$ (°C) | *60.3±0.6* | *47.7±0.6* | *43.0±0.0* | *44.0±1.7* |
|  | $\Delta T_m$ ($WT_{G2}$-mutant) |  | *12.6* | *17.3* | *16.3* |
| **+ Nb11** | $T_m$ (°C) | *577±0.6* | *54.6±0.6* | *53.0±1.0* | *53.7±0.6* |

| | ΔT$_m$ (WT$_{G2}$-mutant) | | 3.1 | 4.7 | 4.0 |

*Table 3: Quantitative evaluation of the effect of Nb11 on the thermal stability of G2 domains. Apparent T$_m$ values from thermofluor experiments were calculated as the minimum of the first derivative of the traces in Figure 6B. Reported values are the average of three independent measurements ± SD, at most. T$_m$ for Nb11 alone in the same experimental conditions is 50 ± 1 °C.*

To quantify the stabilizing effect of Nb11, we performed thermofluor experiments of the G2 domains, whose denaturation is achieved in the presence of a fluorogenic probe (Figure 6B). As reported in Table 3, the T$_m$ values for the mutated G2 domains were lower than that of the WT$_{G2}$ with differences between the T$_m$ value of WT$_{G2}$ and that of the mutated proteins (ΔT$_m$) ranging from 12.6°C to 17.3°C. These results are similar to those previously reported, measured by CD (see supplementary Table 1) [10,14,30,31]. The WT$_{G2}$:Nb11 complex unfolds at a lower temperature than the WT$_{G2}$ alone, suggesting the absence of any stabilization induced by Nb11. On the contrary, mutants' thermal stability significantly benefits from Nb11 binding as indicated by the ΔT$_m$ values which dropped to 3.1-4.7 °C (Table 3).

In agreement with the structural observations, these data show that Nb11 protects G2 from aberrant proteolysis reverting the destabilization induced by the mutations. It is noteworthy that both T$_m$ experimental approaches are insensible to the displacement of the C-terminal tail, because it neither results in a loss of secondary structures nor it leads to the exposure of hydrophobic patches, being the calcium pocket very polar.

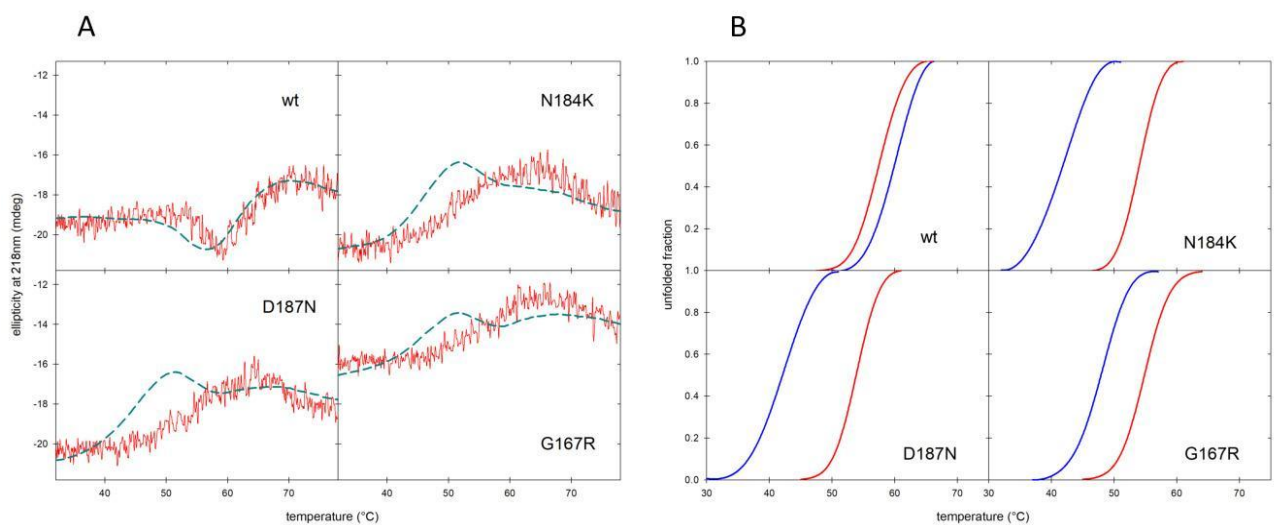

*Figure 6: Thermodynamic stabilities of mutated G2s in the presence or absence of Nb11. A) Unfolding profiles, followed by CD spectroscopy, of WT$_{G2}$, D187N$_{G2}$, N184K$_{G2}$ and G167R$_{G2}$ in complex with Nb11 (red trace) is compared with the arithmetical sum of Nb11 alone + individual G2s (dashed cyan line). B) G2s alone (blue trace) and G2s in complex with Nb11 (red) are analyzed in thermofluor experiments. Thermal stability is measured by SYPRO orange fluorescence, which yields sigmoidal curves typical of a two-state, cooperative denaturation.*

## 3.6 Nb11 counteracts the proteotoxicity of G2 domains in *C. elegans*

It is well known that the administration to nematodes of a toxic compound, such as a misfolded protein, induces a pharyngeal dysfunction [24,25,28,29] that can be evaluated by counting the number of worm's pharyngeal contractions, defined as "pumping rate" [28]. Therefore, we employed *C. elegans* assays to determine whether the structural changes and the stabilizing effects induced by Nb11 on the mutated G2 variants can also affect their biological properties.

First, we evaluated the ability of $WT_{G2}$ or mutated G2 to induce a pharyngeal dysfunction in worms. To this end, 250 μg/ml of each protein was administered to worms, and the pharyngeal contraction was determined at different times after the treatment. After 2 h the pumping rate of $WT_{G2}$-fed worms is significantly reduced compared to those fed vehicle (228.4 ± 1.7 and 217.1 ± 1.6 pumps/min for vehicle- and $WT_{G2}$-fed worms, respectively) (Figure 7A). A greater inhibition was observed in nematodes treated with mutated G2 domains, particularly $N184K_{G2}$ (177.9 ± 1.8 pumps/min) which is significantly more toxic than $G167R_{G2}$ and $D187N_{G2}$ variants (191.6 ± 2.3 and 208.4 ± 3.0 pumps/min, respectively) (Figure 7A). These outcomes were comparable to those of worms treated with 10 mM hydrogen peroxide, used as positive stress control (183.7 ± 1.6 pumps/min, p<0.0001 *vs* vehicle, Student's t-test), indicating that the reduction of pharyngeal contraction induced by the mutated G2 domains is biologically relevant.

We also observe that the pharyngeal impairment induced by $WT_{G2}$ is transient because the pumping rate scored 24 h after the administration was similar to that of vehicle-fed nematodes (233.4 ± 2.3 and 225.8 ± 1.7 pumps/min for vehicle and $WT_{G2}$, respectively). On the contrary, the mutated G2 domains induced a persistent pharyngeal functional damage, without complete recovery even after 24 h (Figure 7B). Whereas the pharyngeal dysfunction induced by $D187N_{G2}$ did not change over time (208.4 ± 3.8 and 211.3 ± 3.5 pumps/min at 2 and 24 h, respectively), those caused by $G167R_{G2}$ and $N184K_{G2}$ significantly decreased (Supplementary Figure 4).

Trying to establish a relationship between protein toxicity and folding, worms were fed with equimolar concentration of either $WT_{G2}$ or $WT_{\Delta G2}$. Lacking the β1 strand, $WT_{\Delta G2}$ is unable to properly fold and displays spectroscopy characteristics typical of a natively unfolded protein under physiological conditions [31]. We observed that the unfolded G2 domain ($WT_{\Delta G2}$) was significantly more effective in reducing the pharyngeal pumping than the folded one underlining the relevance of the folding status for the toxicity (Supplementary Figure 5).

The different toxic potential of the mutated G2 domains is further proven by the experiments evaluating their dose-dependent effects on the pharynx of worms (Figure 7C-D). $IC_{50}$ (half-maximal effect) values calculated 2h after the exposure resulted significantly higher for $WT_{G2}$ than for mutated G2 and, among these, the higher toxicity was obtained for $N184K_{G2}$ (Table 4). Similar $IC_{50}$ values are calculated after 24 h for $G167R_{G2}$ and $D187N_{G2}$ whereas a 4-fold increase is observed for $N184K_{G2}$ (Table 4).

These findings indicate that *C. elegans* can efficiently recognize the proteotoxic potential of G2 domains, pointing to its use as a rapid and valuable tool to investigate the mechanisms underlying AGel. Also, the different proteotoxic abilities exerted by the various mutated G2 domains suggests that each mutation may drive a specific damage.

The effect of Nb11 on the pharyngeal toxicity induced by the G2 domains was then investigated. For these experiments, worms were fed 250 µg/ml of $WT_{G2}$ or $D187N_{G2}$ (corresponding to 19 µM), 100 µg/ml of $N184K_{G2}$ or $G167R_{G2}$ (corresponding to 8 µM), concentrations close to the $IC_{50}$ value calculated for the 2h of exposure (Table 3). G2s were administered alone or as complexes with an equimolar concentration of Nb11. Although Nb11 alone caused a significant reduction of the pharyngeal function, both its toxicity as well as those of G2 domains scored 2 h after administration were neutralized when co-administered (Figure 7E). Noteworthy, the effect of Nb11 on the toxicity induced by the mutated G2 domains were still present 24 h after the treatment (Supplementary Figure 6) as an indication of sustained protective action.

These data indicate that the structural and molecular effects induced by Nb11 on the G2 variants can translate into a biologically relevant effect *in vivo*. Nb11 can bind and chaperone the $D187N_{G2}$ as well as the G2 domains of the renal variants counteracting their proteotoxic potential.

|  | $IC_{50}$ (µg/ml) ± SE | |
|---|---|---|
| Treatment | 2 hours | 24 hours |
| $WT_{G2}$ | 398.5 ± 1.12 | Not interpolated |
| $D187N_{G2}$ | 273.0 ± 1.09 * | 241.8 ± 1.24 |
| $N184K_{G2}$ | 79.7 ± 1.12 **** | 334.5 ± 1.19 ° |
| $G167R_{G2}$ | 133.9 ± 1.20 *** | 170.2 ± 1.21 |

*Table 4: Pumping rates and $IC_{50}$ values of WT and mutated G2 domains. $IC_{50}$ values were calculated from the dose-response curves (Figure 7C/D). * p<0.05, ***p<0.0005 and **** p<0.0001 vs. WT, ° p<0.05 vs. G167R, one-way ANOVA and Bonferroni post-hoc test.*

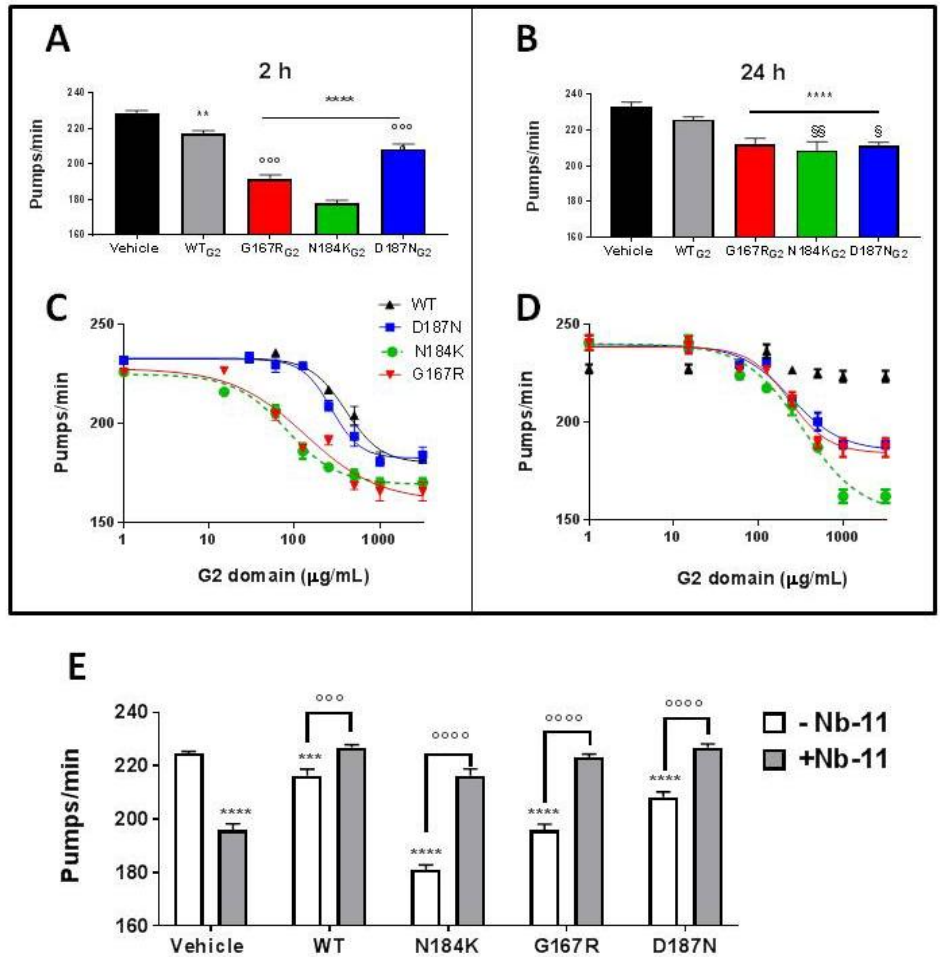

*Figure 7. Proteotoxic effect of the G2 domains and the protective effect of Nb11. Worms fed for 2 h vehicle (2 mM Hepes solution containing 1 mM NaCl and 0.1 mM $CaCl_2$, pH 7.4) or 250 µg/ml $WT_{G2}$, $D187N_{G2}$, $G167R_{G2}$ or $N184K_{G2}$. Pharyngeal pumping was scored (A) 2h and (B) 24h after treatment by counting the number of pharyngeal contractions in one minute (pumps/minute). Data are mean ± SE (N=30 worms/group). \*\* p<0.001 and \*\*\*\* p<0.0001 vs. Vehicle, § p<0.05 and §§ p<0.001 vs. WT, °°° p<0.0005 and °°°° p<0.0001 vs. N184K, one-way ANOVA and Bonferroni's post hoc test (C-D) Dose-response effect of 1-1000 µg/ml of the different G2 domains scored (C) 2h and (D) 24h after the administration. E) G2 domains (250 µg/ml of WT and D187N or 100 µg/ml of N184K and G167R, corresponding to 19 or 8 µM respectively) were incubated with an equimolar concentration of Nb11 (19 or 8 µM) for 10 min at room temperature under shaking conditions. Nb11 (19 or 8 µM) and G2 domains (100-250 µg/ml) alone were also administered. Vehicle (2 mM Hepes solution containing 1 mM NaCl and 0.1 mM $CaCl_2$, pH 7.4) was used as negative control. The solutions were then administered to C. elegans (100 µl/100 worms). The pharyngeal activity was determined 2h after treatment by determining the number of pharyngeal bulb contraction (pumps/min). Data are mean ± SE (N=30 worms/group). \*\*\* p<0.0005 and \*\*\*\*p<0.0001 vs Vehicle without Nb11, °°°° p<0.001 and °°° p<0.005 vs the corresponding sample without Nb11, according to one-way ANOVA and Bonferroni's post hoc analysis. Nb11 treatment significantly protected from the toxicity induced by all WT and mutated G2 domains (interaction p<0.001, two-way ANOVA and Bonferroni's post hoc analysis).*

# 4. Discussion

The current model of the pathological mechanism underlying AGel amyloidosis is mostly based on studies performed on the D187N-mutated protein and experiments conducted *in vitro* and *in vivo* on transgenic animals expressing this variant [15,16,48]. Results obtained with D187N likely also apply to the later discovered Danish variant (D187Y), although some differences were reported [49]. On the contrary, the mechanism underlying the recently discovered renal disease associated with the N184K and G167R mutations have not been fully elucidated. Interestingly, while the crystallographic structures of isolated G2 domains carrying the N184K or G167R mutation are already available [30,31], until now the D187N/Y structural characterization had not been elucidated.

We here demonstrated that crystallization of the D187N G2 domain was possible only in complex with a previously developed nanobody, called Nb11 [19]. Nb11 tightly binds to gelsolin and protects the D187N variant from aberrant furin proteolysis. In the analysis of the crystal structure, we must consider that the binding of Nb11 somehow biases our model. Indeed, we are observing a proteolysis-resistant species that lost, to some extent, the structural determinants of its proteotoxicity.

To elucidate the mechanism of protection exerted by Nb11 it is necessary to understand the pathological mechanism of D187N mutation (and *vice versa*). Clearly, Nb11 acts as a protective chaperone; however, the molecular mechanism behind such function was as yet unclear, mainly because Nb11 binds G2 in a position distant from the furin cleavage site (Figure 2). Regarding the D187N mutation, a large body of literature is already available, and these studies converged to a general agreement, i.e. that the D187N substitution disrupts calcium binding in G2 and the thermodynamic stability of the mutant is decreased to levels similar to those of the WT protein deprived of calcium [10–12,49]. Ultimately the mutation leads to the exposure of an otherwise buried sequence, which is aberrantly cleaved by furin. However, the correlation between calcium binding impairment and susceptibility to proteolysis has been the object of discussion. One hypothesis is that the mutation somehow induces a conformational change of the native state that leads to the exposure of the furin site. Another possibility is that the loss of coordinated calcium increases the population of (partially) unfolded protein, which is generally prone to proteolysis [10–12,49]. The partial disorder hypothesis is consistent with the crystallographic structures, the thermodynamic data as well as with the MD results, which in the absence of the coordinated cation show a fast (on a timescale of tens of ns) opening of the C-terminal stretch. Calcium-mediated disorder-to-order induction has been shown, e.g. in sortase [50,51], adenylate cyclase toxin [52], and possibly several others [53,54].

Interestingly, Kazmirsky and coworkers came to a similar conclusion through a detailed NMR analysis of the D187N variant [12]. In the study, they suggested that the C-terminal tail of the pathogenic variant might be less structured compared to the WT. It might be tempting to explain the increased susceptibility to proteolysis solely based on the flexibility of the C-terminus, which indeed interacts with the hinge loop and exerts some steric protection. However, even the crystallographic structures of the D187N$_{G2}$:Nb11 complex reveal a destabilized tail, and the simulations show a similar dynamic behavior of the stretch irrespective of Nb11 binding. Destabilization of the C-terminus might likely be a necessary condition, but it seems not to be sufficient for the efficient proteolytic cleavage.

In conclusion, we believe that the destabilization of the C-terminal tail is the direct consequence of the loss of the Ca$^{2+}$ ion induced by D187N mutation. Such increased flexibility is the first event that triggers the sequential opening of the gelsolin fold, which results in the exposure of residues of the hydrophobic core. The process is likely reversible and this partially unfolded state of the protein is susceptible to furin proteolysis. In the complex with Nb11, the binding interface, including α0 and α2-loop-β5, becomes more rigidly anchored to the proper WT conformation. The latter Nb11-stabilized region lies at the base of the disorder-prone C terminus, forming an *elbow* of sorts. We speculate that Nb11 stabilization reverses, at least in part, the entropy gain due to the abolished coordination of $Ca^{2+}$.

Even though Nb11 was raised against WT gelsolin, it was shown to bind the D187N variant as well, with a slightly lower affinity [19]. Crystallographic structures of N184K and G167R variants, as isolated G2 as well, are also available [30,31]. N184K neither impairs calcium binding nor it significantly destabilizes the C-terminal tail. Loss of conformational stability of N184K-mutated G2 comes from the rearrangement of the polar contacts, which the mutated residue is part of, in the core of the domain. As for the D187N/Y case, N184K mutation eventually leads to furin proteolysis. Contrarily, the pathological mechanism underlying the G167R-dependent disease is still to be fully elucidated. An alternative amyloidogenic pathway has in fact been proposed based on the observation that this variant dimerizes *via* a domain-swap mechanism [31]. At the same time, even the G167R mutant is prone to aberrant proteolysis and shows impaired thermal stability.

The $WT_{G2}$ and D187N- and N184K-mutated variants share a high structural conservation. The Nb11 binding interface of the $G167R_{G2}$ variant is not affected by its dimerization. As a consequence, it comes as no surprise that Nb11 binds the renal variants with similar efficiency. On the contrary, the protection from proteolysis of the N184K and G167R mutants would have been difficult to predict because the impact of these substitutions on the G2 dynamics as well as the mechanism of destabilization is significantly different.

We tested Nb11 protection of the renal variants in standard furin assays and observed indirect inhibition of the protease activity comparable to that of the D187N variant. To further investigate this aspect, we aimed at testing Nb11's activity on mutated G2 in a more biological context. As neither an animal nor a cell model of renal AGel amyloidosis is currently available, we employed a nematode-based assay already developed and validated by our group and already successfully used to determine the toxicity of different amyloidogenic proteins and investigate the mechanisms underlying their proteotoxic effect [24–27,29]. This is based on the knowledge that the rhythmic contraction and relaxation of the *C. elegans* pharynx, the organ fundamental for the worm's feeding and survival, is sensitive to molecules that can act as "chemical stressors" such as some proteins induced in stress conditions. We here showed for the first time that all isolated G2, at concentrations similar to those observed as circulating in humans [55] can cause a biologically relevant toxic effect recognized by *C. elegans*. Whereas $WT_{G2}$ only induced a transient reduction of the pharyngeal pumping, all the pathological mutants caused an impairment which persisted still after 24 hours, suggesting that these proteins caused a permanent tissue damage.

Interestingly, Nb11, which was able to bind and chaperone all the disease-causing mutant forms of GSN and prevent their aberrant proteolysis, completely abolished the proteotoxic effects induced by G2 domains in worms. Although the mechanisms underlying the ability of *C. elegans* to recognize G2 domain as toxic remains to be elucidated, these findings are consistent with the hypothesis that the toxicity can be ascribed to the partially unfolded status of the proteins and that Nb11 hides the structural determinant of toxicity. The involvement of the folding status in G2 toxicity was further assessed by using a protein totally unfolded due to the truncation of residues 151-167. The toxic effect caused by $WT_{\Delta G2}$ protein on the pharynx of worms was greater and more persistent than that observed with the folded $WT_{G2}$. Additional studies are required to fully elucidate the mechanisms underlying the toxicity of G2 domains on the *C. elegans* pharynx.

The knowledge of high-resolution determinants of D187N toxicity and the therapeutic action of Nb11 may contribute to the ongoing efforts in the expansion of the currently limited landscape of therapeutic interventions against FAF and other amyloidosis-related diseases. The use of the *C. elegans*-based assay for the evaluation of the proteotoxic potential of the G2 domains offers unprecedented opportunities to investigate the molecular mechanisms underlying the AGel forms caused by different GSN variants. This model can also be employed for rapidly screen the protective effect of novel or repurposed drugs thus accelerating the identification of an effective therapy against AGel amyloidosis.

# 5. Conclusions

Nb11 works as a pharmacological chaperone (sometimes referred to as pharmacoperone), a concept originally formulated in the context of small chemicals, rather than macromolecules. Such chaperones restore mutated proteins' function or prevent misfolding, either stabilizing the native fold, or assisting the folding process. Although the pharmacoperone concept has been around for a while, only a limited number of successful examples can be found in literature [56–58], and very few drugs belonging to this family are available on the market. Nb11 is, in our knowledge, the first case of a pharmacoperone-like nanobody whose efficacy has been proven *in vivo*. Contrary to other previously developed pharmacological chaperones, which are mutation-specific, Nb11 protects all gelsolin pathological variants identified so far, irrespective of the different pathways or mechanisms leading to their degradation and/or aggregation. Furthermore, disease-causing mutations yield proteins which are at times difficult to express and purify, due to their solubility, instability or toxicity. Our data show that pharmacological chaperones can be developed using the WT protein as a target, aiming at the stabilization of the native-like conformation, thus increasing the number of diseases that can be potentially tackled with an analogous strategy. Wide-spectrum applicability is of pivotal importance in view of the potential application of Nbs in therapy.

# Acknowledgements

This research was supported by the Amyloidosis Foundation thanks to a Research Grant to MdR. *C. elegans* and OP50 *E. coli* were provided by the *Caenorhabditis* Genetics Center (CGC) which is funded by NIH Office Research Infrastructure Programs (P40 OD010440). TG acknowledges funding from Acellera Ltd. and Università degli Studi dell'Insubria, Italy. This work was partially supported by Fondazione Sacchetti (Grant 2017-2018). The diffraction experiments were performed on beamline ID23-1 at the European Synchrotron Radiation Facility (ESRF), Grenoble, France. We are grateful to local contact at the ESRF for providing assistance in using their beamline.

# Financial & competing interests' disclosure

Prof. Dr. Jan Gettemans has a financial and/or business interests in Gulliver Biomed BVBA, a company that may be affected by the research reported in the enclosed paper.

# Author contribution statement

**TG, JG, LD and MdR**: conceptualization, methodology, supervision. **TG, AB, MM, EM, AV, JG, LD and MdR**: formal analysis, writing – original draft. **TG, DM, AH, MM, AB, MMB** and **MdR:** investigation. **All authors:** writing – review and editing.